%% file: HIG-12-004_temp.tex
\pdfoutput=1

\documentclass[11pt,twoside,a4paper,cmspaper,final,collab]{cms-tdr}
\def\svnVersion{158078}\def\cmsCernNo{2012-176}\def\cmsCernDate{\today}\def\cmsMessage{Submitted to Physical Review Letters}

\begin{document}\cmsNoteHeader{HIG-12-004}

\hyphenation{had-ron-i-za-tion}
\hyphenation{cal-or-i-me-ter}
\hyphenation{de-vices}

\RCS$Revision: 142152 $
\RCS$HeadURL: svn+ssh://alverson@svn.cern.ch/reps/tdr2/papers/HIG-12-004/trunk/HIG-12-004.tex $
\RCS$Id: HIG-12-004.tex 142152 2012-08-02 21:25:30Z chertok $
\newlength\cmsFigWidth
\ifthenelse{\boolean{cms@external}}{\setlength\cmsFigWidth{0.95\columnwidth}}{\setlength\cmsFigWidth{0.75\textwidth}}
\ifthenelse{\boolean{cms@external}}{\providecommand{\cmsLeft}{top}}{\providecommand{\cmsLeft}{left}}
\ifthenelse{\boolean{cms@external}}{\providecommand{\cmsRight}{bottom}}{\providecommand{\cmsRight}{right}}
\providecommand{\PSa}{\ensuremath{a\xspace}}
\providecommand{\PSh}{\ensuremath{\cmsSymbolFace{h}\xspace}}
\providecommand{\PSHp}{\ensuremath{\cmsSymbolFace{H}^+\xspace}}
\providecommand{\PSHm}{\ensuremath{\cmsSymbolFace{H}^-\xspace}}
\providecommand{\PSA}{\ensuremath{\cmsSymbolFace{A}\xspace}}
\cmsNoteHeader{HIG-12-004} 
\title{Search for a light pseudoscalar Higgs boson in the dimuon decay channel in \texorpdfstring{$\Pp\Pp$ collisions at $\sqrt{s} = 7\TeV$}{pp collisions at sqrt(s) = 7 TeV}}

\author{Maria A. Borgia, Maxwell Chertok, John F. Gunion, Mario Pelliccioni}

\date{\today}

\abstract{The dimuon invariant mass spectrum is searched
in the range between 5.5 and 14\GeV for a light pseudoscalar Higgs boson $\PSa$,
predicted in a number of new physics models, including the next-to-minimal supersymmetric
standard model.  The data sample used in the search corresponds to an integrated luminosity
of  1.3\fbinv collected in $\Pp\Pp$ collisions at $\sqrt{s}$ = 7\TeV with the CMS detector
  at the LHC.  No excess is observed above the background predictions and upper limits
  are set on the cross section times branching fraction $\sigma \cdot\mathcal{B}(\Pp\Pp\rightarrow \PSa \rightarrow\Pgmp\Pgmm)$ in the range of 1.5--7.5\unit{pb}.  These results improve on existing  bounds on the $\PSa \bbbar$ coupling
  for $m_{\PSa}  < m_{\PgUa}$ and are the first significant limits for $m_{\PSa} > m_{\PgUc}$.
Constraints on the supersymmetric parameter space are presented in the context of the next-to-minimal model.
}

\hypersetup{%
pdfauthor={CMS Collaboration},%
pdftitle={Search for a light pseudoscalar Higgs boson in the dimuon decay channel in pp collisions at sqrt(s) = 7 TeV},%
pdfsubject={CMS},%
pdfkeywords={CMS, physics, muons, pseudoscalar, higgs}}

\maketitle 

Low energy supersymmetry (SUSY) is an elegant solution to the
hierarchy problem that arises in the Standard Model (SM),
provides a candidate for dark matter, and allows for the
unification of gauge couplings at the grand-unified-theory (GUT) scale~\cite{ref:SUSY-1, ref:SUSY0, ref:SUSY1, ref:SUSY2, ref:SUSY3}.
  However, the minimal supersymmetric model (MSSM) has an ad hoc Higgs superfield mixing parameter ($\Pgm$)  and requires very large masses for the supersymmetric partner of the top quark (stop) in order for the lightest CP-even Higgs boson to be heavier than 122\GeV without large stop mixing \cite{Nakamura:2010zzi}. Both problems are solved
in the next-to MSSM (NMSSM) (a review can be found in Ref.~\cite{NMSSM_rev}),
 which extends the MSSM by introducing a complex singlet
superfield which necessarily contains a scalar field component.  Associated super- and scalar-potential terms generate an effective $\Pgm$ parameter and easily raise the mass of the light Higgs boson without requiring a heavy stop~\cite{nmssmEllisGunion,nmssmEllwanger}.
The added scalar field expands the Higgs sector to three CP-even scalars
($\PSh_1$, $\PSh_2$, $\PSh_3$), two CP-odd scalars ($\PSa_1$, $\PSa_2$) and two
charged scalars (\PSHp, \PSHm). The  $\PSa_1$ is a superposition of the MSSM doublet
pseudoscalar ($\PSa_\mathrm{MSSM}$) and the additional singlet
pseudoscalar of the NMSSM ($\PSa_S$): $\PSa_1 = \cos\theta_\PSA a_\mathrm{MSSM} + \sin\theta_\PSA \PSa_S$,
where $\theta_\PSA$ is the mixing angle.
The NMSSM has two symmetries that, if imposed (e.g. at the GUT scale), imply that small $m_{\PSa_1}$, even $m_{\PSa_1}<2m_\PB$ (where $m_\PB$ is the $\PB$ meson mass), and $|\cos\theta_\PSA|\ll 1$ are very natural possibilities~\cite{nmssmDobrescu}. However, the reduced couplings $C_{\PSa_1 \bbbar} = C_{\PSa_1 \Pgmp\Pgmm} = C_{\PSa_1 \Pgt^+\Pgt^-}=\tanb\cos\theta_\PSA$ can be
sizeable for large values of $\tanb$, the ratio of neutral Higgs field vacuum expectation values, even if $\cos\theta_\PSA$ is small.  More generally, superstring modeling suggests the possibility of many light
$\PSa$ particles, at least some of which couple to $\Pgmp\Pgmm$, $\Pgt^+\Pgt^-$
and $\bbbar$ \cite{Dine:2007zp}. In the following, $\PSa$ ($\PSa_1$) denotes a general (NMSSM) light pseudoscalar
Higgs boson.

Searches for a light $\PSa$ are mainly sensitive to $C_{\PSa\bbbar}$~\cite{nmssmGunion,nmssmLisantiWacker}. For $m_{\PSa}<m_{\PgUa}$, the strongest constraints on $C_{\PSa\bbbar}$ are those from BaBar  \cite{Aubert:2009cp,Aubert:2009cka}.
For $m_{\PSa} > m_{\PgUc}$, only the Tevatron and Large Hadron Collider (LHC) have sensitivity \cite{Dermisek:2009fd}, using production via $\Pg\Pg \rightarrow \PSa$, where the
coupling $C_{\PSa\Pg\Pg}$ derives from quark (especially bottom and top) triangle loops. This process,
plus higher-order corrections, leads to a large cross section due to the large $\Pg\Pg$ parton luminosity at small gluon momentum
fractions, provided the $C_{\PSa\Pq\Paq}$ ($\Pq=\mathrm{t},\mathrm{b}$ in particular) couplings are not too suppressed. This large cross section
will typically lead to a significant number of $\Pg\Pg\rightarrow\PSa\rightarrow\Pgmp\Pgmm$ events even though
$\mathcal{B}(\PSa\rightarrow\Pgmp\Pgmm)$ is small.

In the NMSSM context, where $C_{\PSa_1 \bbbar}=\tanb\cos\theta_\PSA$, the existing limits
  \cite{Apollinari:2005fy,Aaltonen:2009rq}
translate to rather modest limits on $|\cos\theta_\PSA|$.  Such bounds do not strongly constrain
NMSSM models of interest for possibly hiding a light Higgs boson because of $\PSh\to \PSa\PSa$ decays (with $\PSa \to 2\Pgt, 2\Pg, 2\cPqc, 2\cPqs$ decays being dominant \cite{Dermisek:2005ar})
  that are not excluded by large electron-positron (LEP) collider experiments  \cite{Schael:2006cr,ALEPH}.

At tree level, the branching fraction for  $\PSa \rightarrow \Pgmp\Pgmm$
 depends on $m_{\PSa}$ and on $\tan \beta$, but not on
$\cos\theta_\PSA$\cite{Dermisek:2009fd}.  It is nearly constant for
$m_{\PSa} > 5\GeV$ and ranges from $10^{-3}$ to $4 \times 10^{-3}$ for
$\tan\beta = 1$ to $\tan\beta = 50$, changing very little once $\tan \beta > 2$. In contrast,
$\sigma(\Pg\Pg\to \PSa)$ increases rapidly with $\tanb$ due to the fact that
$C_{\PSa \bbbar}\propto \tanb$.  However,
top-quark loop contributions and higher-order corrections imply a slower
$\sigma(\Pg\Pg\to \PSa)$ increase than $\tan^2\beta$. In the context of the NMSSM,
all $\Pq\Paq$ couplings of the $\PSa_1$ are proportional to
$\cos\theta_\PSA$, implying that $\sigma(\Pg\Pg\to \PSa_1)\propto \cos^2\theta_\PSA$.

This Letter presents the results of a search in $\Pp\Pp$ collisions for a light $\PSa$ with a mass
near the $\PgU$ resonances decaying into two oppositely
charged muons.  Data used for this analysis were recorded by the Compact Muon Solenoid (CMS) detector in $\Pp\Pp$ collisions at a center-of-mass energy of 7\TeV, between August and November 2011. The sample corresponds to a total integrated luminosity of 1.3\fbinv, collected with a dedicated trigger.
As estimated in Ref.~\cite{Dermisek:2009fd} and
explicitly demonstrated here, CMS has sensitivity beyond the BaBar and CDF limits,
for the latter due to the higher production yield
[$\sigma_\mathrm{LHC}(\Pp\Pp\rightarrow \PSa) \sim 4.5 \sigma_\text{Tevatron}
(\Pp\Pap\rightarrow \PSa$)] and the higher acceptance and efficiency of the
muon detector.  Furthermore, the CMS analysis can extend the limits into the $m_{\PSa} >m_{\PgUc}$ mass range.

The central feature of the CMS detector is a superconducting solenoid, of 6\unit{m} internal diameter, providing a magnetic field of 3.8\unit{T}. Within the field volume are the silicon pixel and strip tracker, the crystal electromagnetic calorimeter and the brass/scintillator hadron calorimeter. Muons are measured by gas-ionization detectors embedded in the steel return yoke in the pseudorapidity range $|\eta|< 2.4$,
($\eta \equiv - \ln[\tan(\theta/2)]$, where $\theta$ is the polar angle
          of the trajectory of a particle with respect to the direction of
          the counterclockwise proton beam)
using three detector technologies: drift tubes (DT)
(for the range $|\eta|< 1.2$), resistive plate chambers (RPC) (for $|\eta|< 1.6$) and
cathode strip chambers (CSC) (for $0.9 <|\eta|< 2.4$).
The DT and RPC are indicated as the central "barrel" while the CSC comprises the "endcaps".
A more detailed description of the CMS detector can be found in~\cite{CMSDetector2008}.

We search the dimuon invariant mass distribution between 5.5 and 8.8\GeV (defined as ``mass range 1'') and between 11.5 and 14\GeV (``mass range 2'') for a narrow  resonance $\PSa$, with a decay width ${\sim}\!\MeV$,  natural in the NMSSM context.
We avoid the range between 9 and 11\GeV because the abundant contributions of the bottomonium resonances to the mass spectrum makes this search unfeasible.  Selection criteria are applied to reduce backgrounds from QCD continuum,
and we perform a mass scan in mass ranges 1 and 2 to determine a potential contribution from an $\PSa$ signal.  Given the better mass resolution in the barrel part of the detector than in the endcaps, we also separate the mass scan into two acceptance regions, based on the dimuon $\eta$, in order to improve the sensitivity.

We analyze events collected with an online selection that requires the detection of two opposite-sign muons with transverse momenta $\pt > 3.5\GeV$ and additional requirements imposed at the high level trigger (HLT). All three muon systems, DT, CSC, and RPC, take part in the trigger decision.
A good primary vertex is also required, as defined in Ref.~\cite{CMS-PAS-TRK-10-005}.
The additional HLT requirements  include $\pt(\Pgmp\Pgmm) > 6\GeV$,
$5.5 < m_{\Pgmp\Pgmm} < 14\GeV$,
and a distance of closest approach of the muon tracks to the beam axis compatible with that
expected for prompt decays.
A prescale factor of two was imposed on the trigger to maintain a reasonable trigger rate.

The main backgrounds arise from QCD processes and, in the lower invariant mass range, from a residual tail of the $\PgUa$ resonance. We determine the background shape in the invariant mass directly from data, and use simulated events
as a cross-check. Signal samples,
QCD,  and $\PgU$ resonances are simulated with \PYTHIA 6.4.24, Tune D6T \cite{Sjostrand:2006za}, and CTEQ6 parton distribution functions \cite{Pumplin:2002vw}.  Tune Z2 gives compatible results.  As the NMSSM is not fully implemented in \PYTHIA, we generate the MSSM pseudoscalar $\PSA$ boson in the mass range of 5.5 to 14\GeV and require dimuon decays.
These samples also contain a simulation of the effects on the number of primary vertices from overlapping $\Pp\Pp$ interactions in the same bunch crossing.

To select the best dimuon candidate in each event, quality criteria are applied to the tracks which reject misidentified muons and muons from kaon and pion decays.
Muons are required to be within the geometrical acceptance ($|\eta| \leq$ 2.4) and to be in the plateau of the trigger efficiency, with $\pt > 5.5$\GeV. Muon tracks are required to have at least 11 hits in the silicon tracker, at least one of which must be in the pixel detector, and a track fit $\chi^2/\mathrm{dof} < 1.8$.  This value is chosen to maximize the signal significance with respect to the QCD continuum, which is extracted directly from data.

Isolation requirements  suppress misidentified leptons from jets and non-prompt muons from hadron decays. Muons are required to be isolated within a cone of radius $\Delta R =
\sqrt{(\Delta\eta)^2+(\Delta\phi)^2}$ = 0.3 around the muon direction, where $\phi$
is the azimuthal angle.
The muon isolation, $I_\text{rel}$, is defined as the sum of the \pt\ (as measured in the silicon tracker) and transverse energy (as measured in the calorimeters)  of all objects within this cone (excluding the muon itself), divided by the muon \pt\ as measured by the tracker.  We require $I_\text{rel}<0.2$.
This requirement is optimized by comparing the simulated $\PSa$ signal with opposite-sign dimuons from data, and we verify that this value is appropriate for both the
barrel and endcap dimuon pairs.  This isolation requirement rejects a
large fraction of the background arising from QCD production of jets.

Dimuon candidates consist of two opposite-sign muons \cite{CMS-PAS-MUO-10-002} with an invariant mass between 5.5 and 14\GeV.  If more than one dimuon candidate is present, that with the highest $\chi^2$ probability associated to the kinematic fit of the dimuon vertex is retained.

The invariant mass spectrum in the search range has two main contributions: the QCD continuum and the bottomonium resonances. To characterize these shapes for use in the mass scan, we perform a binned maximum likelihood fit to the total invariant mass distribution. For the QCD continuum, we use a first-order polynomial probability density function (PDF). Each $\PgU$ resonance is parametrized via a double Crystal Ball (CB) function \cite{Oreglia:1980cs,Khachatryan:2010zg}.
The resolution of one of the CB functions is left free in the fit but is constrained to be the same for all the three resonances. The resolution of the second CB function is determined from the fit of the $\PgUa$ peak, and forced to scale with the mass of the other two resonances. As the resonances overlap, we fit for the presence of all three $\PgU$ states simultaneously using three double CB functions. The mean of the CB of the $\PgUa$ is left free in the fit to accommodate a possible bias in the momentum scale calibration. The number of free parameters is reduced by fixing the $\PgUb$ and $\PgUc$ mass differences, relative to $\PgUa$, to their world average values \cite{Nakamura:2010zzi}.

\begin{figure}
  \begin{center}
                    \includegraphics[width=\cmsFigWidth]{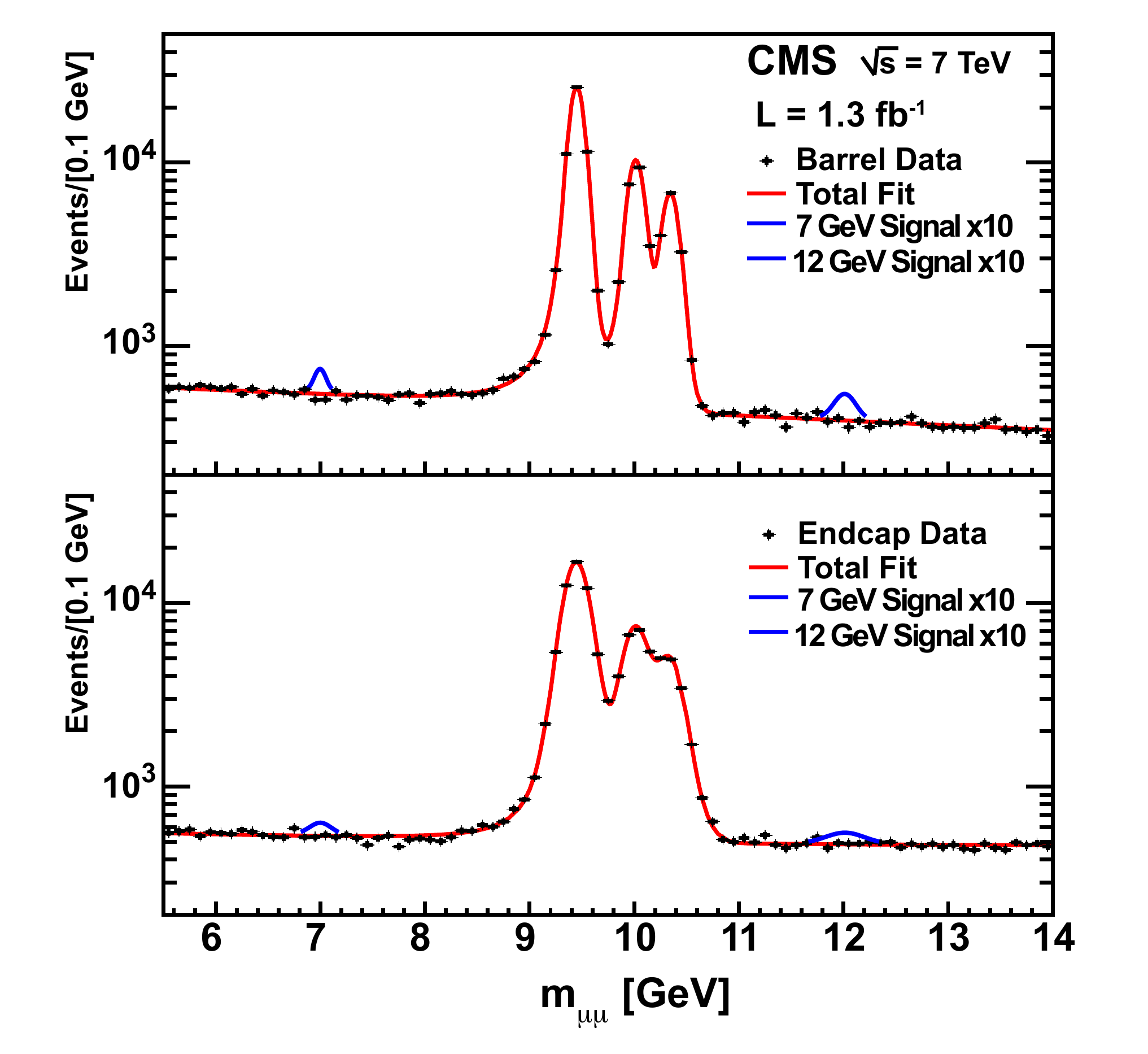}
        \caption{
Dimuon invariant mass distribution for the barrel (upper) and endcaps (lower) after the event selection. The invariant mass distributions are fitted accounting for the three $\PgU$ resonances and QCD continuum.  Hypothetical signals
from pseudoscalar Higgs bosons $\PSa$ at 7 and 12\GeV are shown.
	}
    \label{fig:dimumass}
  \end{center}
\end{figure}

The fits to the $\PgU$ shape and continuum background are performed in the barrel and endcap regions separately, shown in Fig.~\ref{fig:dimumass}.  The fitted numbers of events are given in Table~ \ref{tab:Yfit}; the barrel/endcaps ratio for the $\PgU$ peaks is consistent with MC predictions.      Outside the $\PgU$ peak range, corresponding to the signal search mass ranges, the data are well-described by a first-order polynomial.  Figure~\ref{fig:dimumass} also shows hypothetical signals from pseudoscalar Higgs bosons $\PSa$ with masses 7 and 12\GeV, and 2\unit{pb} cross sections, scaled by a factor of 10 for visibility.
We perform mass scans of the invariant mass spectra, dividing mass range 1 into 110 steps and mass range 2 into 100 steps of 30\MeV each, and treating the barrel and endcap spectra separately. At each step, we build a signal Gaussian PDF with a mean fixed to the center of the step and a width determined by the mass resolution, use a first-order polynomial to characterize the background, and perform an unbinned maximum likelihood fit to search for a possible contribution from the $\PSa$.
For each signal mass point, we determine the resolution by fitting the $\PSa$ invariant mass spectrum with two CB functions (as for the $\PgU$, the sum of two CB functions better describes the resolution) and the mass
resolution is calculated as the weighted average of the widths of the two functions.
The resulting dimuon invariant mass resolution ranges from 50 to 120\MeV (90 to 190\MeV) in the barrel (endcaps) for the mass range 5.5 to 14\GeV.   These agree well with the resolution obtained from the $\PgU$ resonances in data and MC.  We fit the resolution as a function of  mass using the simulated signal samples, and use this to extract the values of the dimuon mass resolution for each mass bin needed in the scan to determine the upper limit.

In mass range 1, we take into account the radiative tail of the $\PgUa$ by including its shape determined from the full invariant mass spectrum fit.  No significant discrepancy with SM background predictions is observed, and we proceed to set cross section limits, as described below.

\begin{table}[htbH]
\begin{center}
\topcaption{Fitted numbers of $\PgU$ and continuum background events in the invariant mass range 5.5--14\GeV. The $\PgU$ contributions are summed over the three resonances.\label{tab:Yfit}}
\begin{tabular}{ccc}
\hline
Contribution         & events (barrel) & events (endcaps)\\
\hline
$\PgU$           &   $93753\pm 396$          &   $95876\pm 454$     \\
Continuum       &   $41210\pm 320$          &   $45792\pm 385$     \\
\hline
\end{tabular}
\end{center}
\end{table}

The efficiency for the selection is factorized into three  contributions, $\epsilon = \epsilon_\text{acc} \times \epsilon_\text{trig} \times \epsilon_\text{sel}$,
where $\epsilon_\text{acc}$ is the kinematic acceptance for the $\PSa$, $\epsilon_\text{trig}$ is the efficiency of the muon trigger, and $\epsilon_\text{sel}$ is the efficiency of the selection applied to the dimuon candidates.
We use \PYTHIA~6 to simulate the $\PSa$ signal and to determine $\epsilon_\text{acc}$.
The trigger and selection efficiencies ($\epsilon_\text{trig}$ and $\epsilon_{sel}$) are measured with $\JPsi$ events in data using the tag-and-probe technique \cite{Khachatryan:2010zg}. We perform this study in bins of $\eta$ and \pt\ of the probe muon. The efficiency values extracted from data are compared with those obtained from the simulation of prompt $\JPsi \to \Pgmp\Pgmm$. The difference between the efficiency in data and Monte Carlo (MC) simulation is evaluated in bins of \pt\ and $\eta$ and used as a correction to weight the MC events in order to accommodate possible  discrepancies. These corrections are typically on the order of a few percent. For each dimuon candidate, the weight is the product of the corrections for the two muons.

The isolation requirement efficiency that contributes to $\epsilon_\text{sel}$ cannot be measured using the $\JPsi$ dataset as one of the main production mechanisms for $\JPsi$ is through \PB-meson decays, resulting in non-isolated muons. This is not well accounted for in simulation, and would result in biased data/MC efficiency corrections. In order to estimate this correction, we use $Z\rightarrow\Pgmp\Pgmm$ events and consider the lower \pt\ spectrum of the probe muon.

The total efficiency $\epsilon$ is defined for each $\PSa$ mass sample as the fraction of generated signal events, weighted by the appropriate data/MC corrections, that satisfy all the selection requirements.  This ranges from 1--3.5\% for the $\PSa$ mass
range of 5.5--14\GeV, and we fit the $\epsilon$ distributions with second (third) order polynomial functions in the barrel (endcaps) to use in the mass scan. The increase in the efficiency as a function of the invariant mass is mainly due to the \pt\ requirements on the muons at the HLT level.

Several sources of systematic uncertainty affect these results, including a 2.2\% uncertainty on the integrated luminosity \cite{CMS-PAS-SMP-12-008}.
The efficiency corrections are determined using the tag-and-probe results described above.
We determine, event-by-event, the uncertainty on the total efficiency corrections  by propagating the uncertainties on the single muon corrections.
This total event efficiency uncertainty is largely independent of mass, with a maximum value of  12\%. We apply this value as a systematic uncertainty for every bin in the scan.

The isolation efficiency is uncertain at the 5\% level, corresponding to the largest discrepancy between data and MC simulation in the entire relevant \pt\ range.
We evaluate the systematic uncertainty on the resolution of the $\PSa$ as the quadrature sum of the difference between the mass resolution of the $\PSa$ with a mass of 10\GeV\ and the resolution of the $\PgUb$ (which has the same mass) in MC simulation, and the difference between the latter and the mass resolution obtained for the $\PgUb$ from data.
Additionally, the finite statistics for the determination of the mass resolution as a function of the dimuon mass contributes a source of uncertainty.  We consider the mass ranges separately and  include these systematic uncertainties in the calculation of the upper limit on the cross section times branching fraction.  Overall, this adds a 11\% (4\%) effect for the barrel (endcaps).

Systematic uncertainties on the background description include the shape uncertainty of the first-order polynomial fit of the background PDF.  We fit the background with alternative functions (a second-order polynomial and an exponential function), generate MC pseudo-experiments using these functions, and fit the distributions using the first-order polynomial.  The resulting systematic uncertainties, from the distribution of the fitted parameters, is of the order of a few percent.

\begin{figure}[hbtp]
  \begin{center}
    \includegraphics[width=\cmsFigWidth]{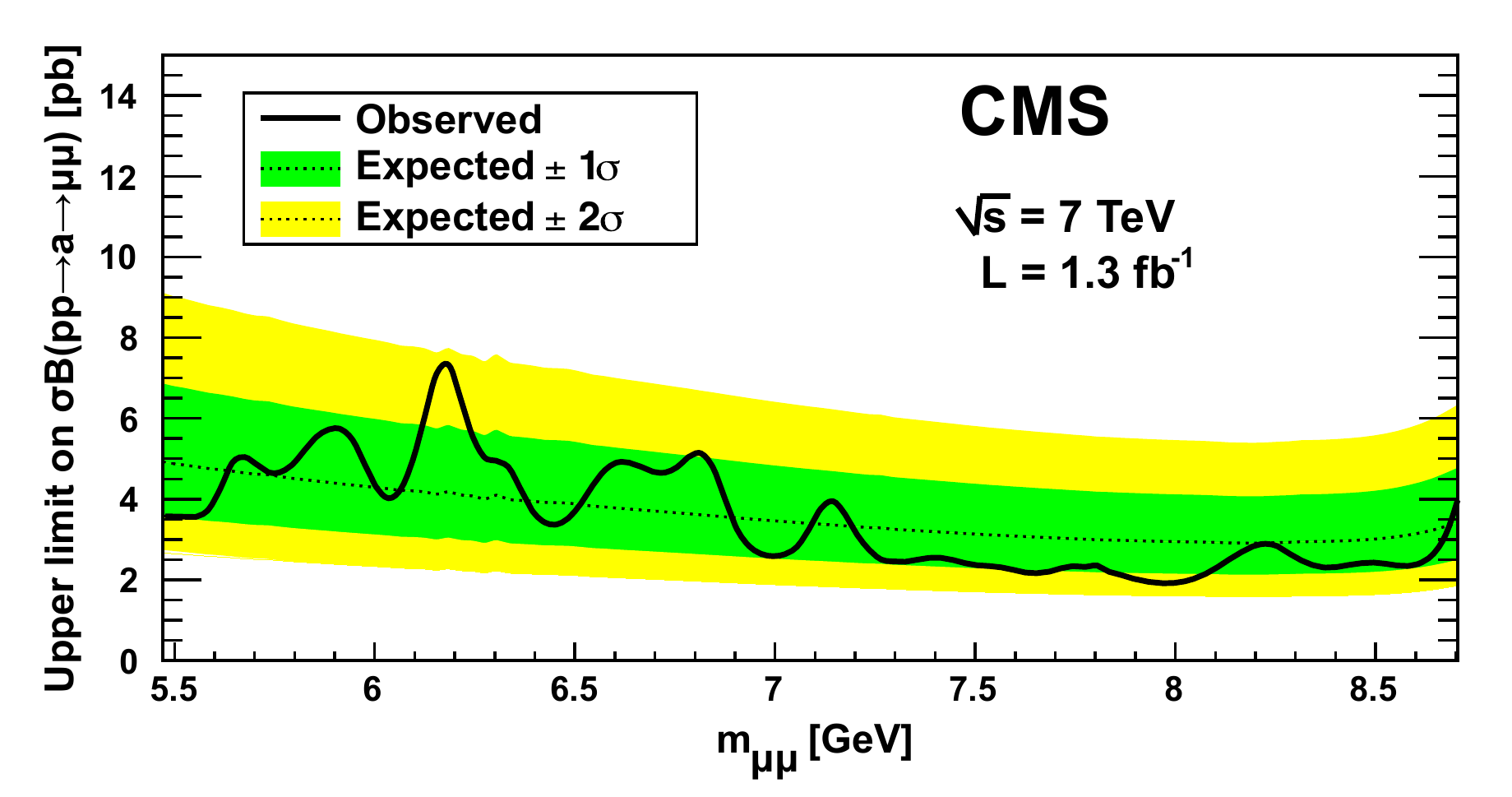}
      \includegraphics[width=\cmsFigWidth]{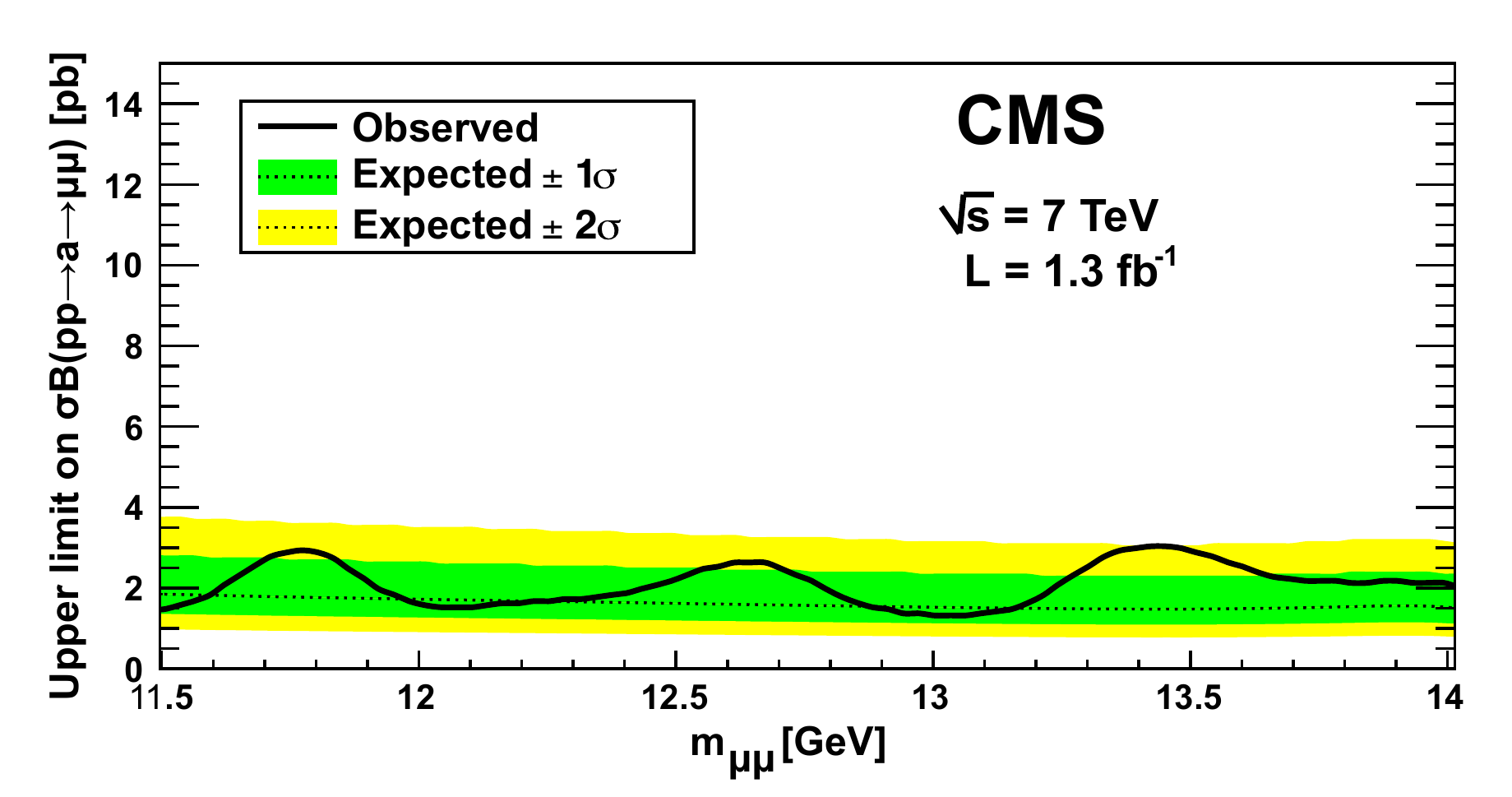}
     \caption{Upper limits at 95\% CL on $\sigma \cdot \mathcal{B}(\Pp\Pp\rightarrow a\rightarrow\Pgmp\Pgmm)$ in mass range 1 (upper panel) and mass range 2 (lower panel) including systematic uncertainties.  The dotted lines correspond to the expected limits, and the bands correspond to 1- and 2-$\sigma$ level uncertainties on the expected limits.}
    \label{fig:UL}
  \end{center}
\end{figure}

\begin{figure}[hbtp]
  \begin{center}
            \includegraphics[width=\cmsFigWidth]{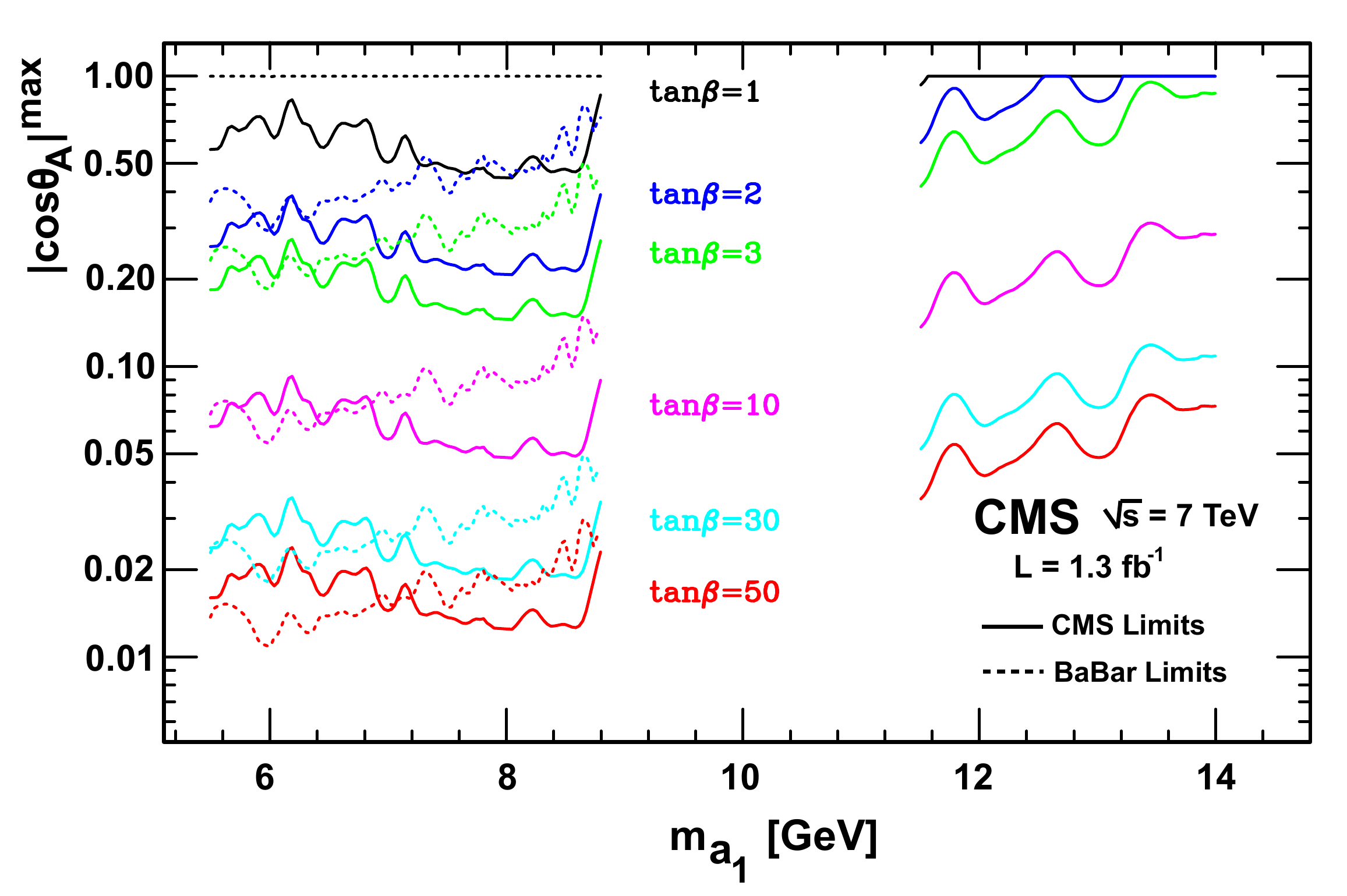}
     \caption{Upper limits on the NMSSM parameter $|\cos\theta_\PSA|$ as a function of $m_{a_1}$ in the two mass ranges.  The solid curves correspond to different $\tanb$ values:
     from top to bottom,
      $\tanb=1$,
      $\tanb=2$, $\tanb=3$,
    $\tanb=10$, $\tanb=30$, and $\tanb=50$. For each $\tanb$ value in mass range 1 the second,
    dotted curve shows the limits from the BaBar $\PgU$ analysis.  There are no BaBar
    limits for $\tan\beta=1$ in mass range 1, or for any $\tan\beta$ in mass range 2.
    The line at $\abs{\cos\theta_\PSA}^\text{max}=1$ is
    equivalent to no limit.  Results from CDF are not shown as they are less stringent than the BaBar limits.}
    \label{fig:ctamax}
  \end{center}
\end{figure}

No significant signal is observed, and
we determine $95\%$ confidence level (CL) upper limits on $\sigma \cdot \mathcal{B}(\Pp\Pp\rightarrow a\rightarrow\Pgmp\Pgmm)$ as a function of the dimuon mass using the CLs approach~\cite{Junk:1999kv,Read:2002hq,LHC-HCG-Report}.
A few steps at the edges of the mass scans, where the fitting procedure has no predictive power on the signal shape, are not used.
Figure \ref{fig:UL} shows the upper limit results for the two mass ranges including the systematic uncertainties discussed above.
These limits are significant in the context of the
NMSSM, and can be presented in terms of upper
limits on $|\cos\theta_\PSA|$.  The larger the value of $\tanb$,  the stronger is the
constraint.
Figure \ref{fig:ctamax} presents upper limits, $|\cos\theta_\PSA|^\text{max}$
as a function of $m_{a_1}$ for $\tanb=1,2,3,10,30,50$.
Our upper limits are compared to an earlier analysis of the BaBar $\PgUa$ and $\PgUc$ data \cite{a_excl1}, and are superior for $m_{a_1}\ge 7.5\GeV$ for $\tan\beta = 50$,
decreasing to $m_{a_1}\ge 6\GeV$ for $\tan\beta = 2$, and are superior for
all masses at $\tan\beta=1$.  Further, these are the first significant limits for $m_{\PSa} > m_{\PgUc}$.

In conclusion,
we performed a search for a narrow, low mass pseudoscalar $\PSa$, which
is produced by $\Pg\Pg\to \PSa$ and decays via $\PSa\rightarrow\Pgmp\Pgmm$ in the
mass ranges 5.5--8.8\GeV and 11.5--14\GeV, using a data sample
corresponding to an integrated luminosity of  1.3\fbinv collected with the CMS detector.
No significant signal is observed, and we
set  upper limits on $\sigma \cdot \mathcal{B}(\Pp\Pp\rightarrow a\rightarrow\Pgmp\Pgmm)$.
These upper limits are applied in the context of the light pseudoscalar $\PSa_1$ of the
NMSSM to yield upper limits on the NMSSM parameter
$\abs{\cos\theta_\PSA}$.
These limits are superior to existing constraints for a significant portion
of the $m_{\PSa_1} < m_{\PgUa}$ mass range, and are the first significant limits
available in the $m_{\PSa_1} > m_{\PgUc}$ mass range.

We congratulate our colleagues in the CERN accelerator departments for the excellent performance of the LHC machine. We thank the technical and administrative staff at CERN and other CMS institutes, and acknowledge support from: FMSR (Austria); FNRS and FWO (Belgium); CNPq, CAPES, FAPERJ, and FAPESP (Brazil); MES (Bulgaria); CERN; CAS, MoST, and NSFC (China); COLCIENCIAS (Colombia); MSES (Croatia); RPF (Cyprus); MoER, SF0690030s09 and ERDF (Estonia); Academy of Finland, MEC, and HIP (Finland); CEA and CNRS/IN2P3 (France); BMBF, DFG, and HGF (Germany); GSRT (Greece); OTKA and NKTH (Hungary); DAE and DST (India); IPM (Iran); SFI (Ireland); INFN (Italy); NRF and WCU (Korea); LAS (Lithuania); CINVESTAV, CONACYT, SEP, and UASLP-FAI (Mexico); MSI (New Zealand); PAEC (Pakistan); MSHE and NSC (Poland); FCT (Portugal); JINR (Armenia, Belarus, Georgia, Ukraine, Uzbekistan); MON, RosAtom, RAS and RFBR (Russia); MSTD (Serbia); MICINN and CPAN (Spain); Swiss Funding Agencies (Switzerland); NSC (Taipei); TUBITAK and TAEK (Turkey); STFC (United Kingdom); DOE and NSF (USA).

\bibliography{auto_generated}   

\cleardoublepage \appendix\section{The CMS Collaboration \label{app:collab}}\begin{sloppypar}\hyphenpenalty=5000\widowpenalty=500\clubpenalty=5000\input{HIG-12-004-authorlist.tex}\end{sloppypar}
\end{document}

%% file: HIG-12-004-authorlist.tex
\textbf{Yerevan Physics Institute,  Yerevan,  Armenia}\\*[0pt]
S.~Chatrchyan, V.~Khachatryan, A.M.~Sirunyan, A.~Tumasyan
\vskip\cmsinstskip
\textbf{Institut f\"{u}r Hochenergiephysik der OeAW,  Wien,  Austria}\\*[0pt]
W.~Adam, T.~Bergauer, M.~Dragicevic, J.~Er\"{o}, C.~Fabjan\cmsAuthorMark{1}, M.~Friedl, R.~Fr\"{u}hwirth\cmsAuthorMark{1}, V.M.~Ghete, J.~Hammer, N.~H\"{o}rmann, J.~Hrubec, M.~Jeitler\cmsAuthorMark{1}, W.~Kiesenhofer, V.~Kn\"{u}nz, M.~Krammer\cmsAuthorMark{1}, D.~Liko, I.~Mikulec, M.~Pernicka$^{\textrm{\dag}}$, B.~Rahbaran, C.~Rohringer, H.~Rohringer, R.~Sch\"{o}fbeck, J.~Strauss, A.~Taurok, P.~Wagner, W.~Waltenberger, G.~Walzel, E.~Widl, C.-E.~Wulz\cmsAuthorMark{1}
% ...
\vskip\cmsinstskip
\dag:~Deceased\\
1:~~Also at Vienna University of Technology, Vienna, Austria\\
% ...
